\documentstyle[prd,preprint,aps]{revtex}

\begin{document}

\draft

\font\large=cmbx10 at 12 pt  
\newcount\equationno      \equationno=0
\newtoks\chapterno \xdef\chapterno{}
\def\eqn{\eqno\eqname}
\def\eqname#1{\global \advance \equationno by 1 \relax
\xdef#1{{\noexpand{\rm}(\chapterno\number\equationno)}}#1}

\def\la{\mathrel{\mathchoice 
{\vcenter{\offinterlineskip\halign{\hfil
$\displaystyle##$\hfil\cr<\cr\sim\cr}}}
{\vcenter{\offinterlineskip
\halign{\hfil$\textstyle##$\hfil\cr<\cr\sim\cr}}}
{\vcenter{\offinterlineskip
\halign{\hfil$\scriptstyle##$\hfil\cr<\cr\sim\cr}}}
{\vcenter{\offinterlineskip
\halign{\hfil$\scriptscriptstyle##$\hfil\cr<\cr\sim\cr}}}}}

\def\ga{\mathrel{\mathchoice 
{\vcenter{\offinterlineskip\halign{\hfil
$\displaystyle##$\hfil\cr>\cr\sim\cr}}}
{\vcenter{\offinterlineskip
\halign{\hfil$\textstyle##$\hfil\cr>\cr\sim\cr}}}
{\vcenter{\offinterlineskip
\halign{\hfil$\scriptstyle##$\hfil\cr>\cr\sim\cr}}}
{\vcenter{\offinterlineskip
\halign{\hfil$\scriptscriptstyle##$\hfil\cr>\cr\sim\cr}}}}}
\def\s{\smallskip}
\def\n{\noindent}

\def\x{{\bf x}}
\def\y{{\bf y}}
\def\R{{\cal R}}

\reversemarginpar


\title{Quantum gravitational corrections to propagator in 
arbitrary spacetimes}
\author{T. Padmanabhan\thanks{paddy@iucaa.ernet.in}}
\address{IUCAA, Post Bag 4, Ganeshkhind, Pune 411 007, India.}
\maketitle

\begin{abstract}
The action for a relativistic free particle of mass $m$ receives a 
contribution $-m\R (x,y)$ from a path of length $\R(x,y)$ connecting 
the events $x^i$ and $y^i$. 
Using this action in a path integral, one can obtain the Feynman propagator 
for a spinless particle of mass $m$ in any background spacetime. 
If one of the effects of quantizing gravity is to introduce a minimum 
length scale $L_P$ in the spacetime, then one would expect the segments 
of paths with lengths less than $L_P$ to be suppressed in the path integral. 
Assuming that the path integral amplitude is invariant under the 
`duality' transformation ${\cal R}\to L_P^2/\R$, one can calculate 
the modified Feynman propagator in an arbitrary background spacetime. 
It turns out that the key feature of this modification is the 
following:~The proper distance $(\Delta x)^2 $ between two events, 
which are infinitesimally separated, is replaced by $\Delta x^2 
+ L_P^2$; that is the spacetime behaves as though it has a `zero-point 
length' of $L_P$. 
This equivalence suggests a deep relationship between introducing a
`zero-point-length' to the spacetime and postulating invariance of 
path integral amplitudes under duality transformations. 
In the Schwinger's proper time description of the propagator, the 
weightage for a path with proper time $s$ becomes $m(s+L_P^2/s)$ 
rather than as $ms$. 
As to be expected, the  ultraviolet behavior of the theory is improved
significantly and divergences will disappear if this modification is 
taken into account. 
Implications of this result are discussed.
\end{abstract}
\pacs{PACS number(s): 04.60. -m, 11.25. -w, 11.25. Sq}

\section{Introduction and Summary}

It has been conjectured for a long time that the spacetime structure
at very small scales [close to  $L_P \equiv (G\hbar/c^3)^{1/2}$] will
be drastically affected by quantum gravitational effects. Since any
quantum field has virtual excitations of arbitrary high energy - which
probe arbitrary small scales - it follows that the conventional quantum
field theory can only be an approximate description, valid at energies
far smaller than Planck energies. 
The `correct' description of nature has to take into account quantum 
nature of spacetime geometry and should reduce to the conventional 
description at low energies. 
Can we say anything about the kind of modifications quantum gravitational
effects will introduce into the description of other quantum fields? 
I investigate some aspects of this question in this paper.

Let us focus attention on a  scalar field $\phi(x)$ of mass $m$ in a
$D$-dimensional Euclidean spacetime. 
Eventually we are interested (probably!) in the case of $D=4$ 
Lorentzian spacetime, which can be achieved by suitable analytic 
continuation. 
Since all matter generates and couples to gravity, there is no such 
thing as a {\it free} scalar field; in the least, one should grant 
the fact that the scalar field is coupled to its own self-gravity. 
So, in general, the action $A[\phi, g_{ik}]$ describing the system 
will be a functional of both $\phi(x)$ {\it and}  the metric $g_{ik}(x)$ 
of the spacetime.
The full quantum field theory of such a system will be based on a formal
path integral like
\begin{equation}
{\cal G}=\sum_{g,\phi}\exp(-A[g,\phi])\label{eqn:orgsum}
\end{equation}
The Feynman propagator $G_F(x,y)$ for the scalar field, (and higher 
order $n$-point functions, all of which can be obtained from a path 
integral description)  will contain information about the quantum 
mechanical properties of $\phi$. 

To the extent we can ignore the gravitational coupling, we can  have a 
free scalar field in flat spacetime and the evaluation of ${\cal G}$ 
is trivial.
At the next level, if we treat the background spacetime 
as curved but classical, one can ignore the sum over 
metrics in equation~(\ref{eqn:orgsum}) and construct the 
propagator $G_F(x,y|g)$ in a given background metric $g_{ik}$. 
We do not have closed form for this in an arbitrary background because 
the partial differential equation for $G_F(x,y|g)$ has no closed form 
solution in an arbitrary background. 
What is more important, {\it such a propagator cannot be trusted when
$(x-y)^2<L_P^2$ since the quantum gravitational fluctuations of the 
background geometry cannot be ignored at these scales} and our 
approximation of working with a fixed background $g_{ik}$ breaks down.
We need to know how the quantum fluctuations of the metric affect the 
propagator $G_F(x,y|g)$ at these scales. 

This is quite a different question from the one usually addressed in the 
subject of quantum fields in curved spacetime in which one worries how 
the quantum nature  of the scalar field affects the background geometry 
(`back reaction'). Such an issue can be tackled, for example, by 
integrating out $\phi$ in equation~(\ref{eqn:orgsum}) and obtaining an 
effective action for gravity, say. 
In contrast, we are interested in how the quantum fluctuations of 
$g_{ik}$ affect the quantum propagator for the scalar field. 
Formally, if we write $g_{ik}=\bar g_{ik}+h_{ik}$, where $\bar g_{ik}$ 
is the, average, large scale spacetime metric and $h_{ik}$ are the small 
scale quantum fluctuations, then we are interested in the effect of 
summing over the fluctuations  $h_{ik}$ in equation~(\ref{eqn:orgsum}) 
to get low-energy quantum theory for the scalar field in the background
metric $g_{ik}$. 
The resulting propagator  $G_F(x,y|\bar g)$, for example, can be 
thought of as the one found by averaging $G_F(x,y|g)$ over the 
quantum  fluctuations in $g_{ik}$ around $\bar g_{ik}$. 
In particular, $\bar g_{ik}$ could just be flat spacetime metric 
$\eta_{ik}$. 
Even in this case, we expect the quantum fluctuations of gravity to 
modify the propagator for $(x -y)^2 < L^2_p$ (or, in momentum space, 
for $(p^2 + m^2) L^2_p > 1$). 
The concept of a {\it free}\/ quantum field is an approximate, 
lower-energy, notion and we do have to change it for $(x - y)^2 < L^2$. 
(In fact, even the description in terms of a field may be inadequate at 
short distances and we may need string theory or models based on Ashtekar
variables.) 
Can we capture the key effects quantum gravitational fluctuations by 
invoking some general principle?

To address this question, it is convenient to write $G_F(x,y|g)$ in an
alternative form. 
We know that the propagator in a given background metric can be expressed 
in two equivalent forms as 
\begin{equation}  
G_F(x,y|g)
=\sum\limits_{\rm paths} e^{-m{\cal R}(x,y)} 
= \int\limits^{\infty}_0 d \tau e^{-m^2 \tau} 
\int {\cal D} x \exp\left(- {1\over 4} 
\int^{\tau}_0 g_{ik} \dot x^i \dot x^k d \eta \right)\label{eqn:equform}
\end{equation}
In the first form, ${\cal R}(x,y|g)$is the proper length of a path 
connecting the events $x$ and $y$, calculated with the background 
metric $g_{ik}$ and the sum is over all paths. 
The action $m{\cal R}$ has a square root in it but can be evaluated 
by standard lattice techniques (see next section).
It is also possible to show by these methods that the result is 
equivalent to the second expression. 
This expression, which is originally due to Schwinger, has a simple 
physical interpretation. 
By rescaling the time variable from $\eta$ to $ s\equiv m\eta$ and 
$\tau $ to $\tau'\equiv m\tau$ we can change the factor $\exp(-m^2\tau)$ 
to $\exp(-m\tau')$ and the path integral Kernel to 
\begin{equation}
K(x,y,\tau' |g)\equiv 
\int {\cal D} x \exp\left(- {m \over 4} 
\int^{\tau'}_0 g_{ik} \dot x^i \dot x^k ds \right)\label{eqn:defker}
\end{equation}
This kernel can be thought of as the probability amplitude for a 
particle to propagate from $x$ to $y$ in a proper time interval 
$\tau'$ in a given background spacetime. 
The amplitude for propagation with energy $E$ (in the rest frame) 
is given by the  Fourier transform of $K(x,y,\tau' |g)$ in the time 
variable $\tau'$, with respect to $E$ in the Lorentzian space; in 
the Euclidean space, it will be a Laplace transform. 
Setting the energy in the rest frame equal to $m$ we obtain the 
expression in equation ~(\ref{eqn:equform}). 
(The physical interpretation of these expressions and their relationship 
to Jacobi action etc are explored in detail in ref.~\cite{paddy94}).

The above expressions assume that we have a  classical background 
spacetime with a given, fixed, metric. As we said before, such a 
description is bound to break down when $(x-y)^2<L_P^2$. 
More generally, equations~(\ref{eqn:equform}), and~(\ref{eqn:defker}) 
sum over paths which probe arbitrarily small scales at which the metric
fluctuations are likely to be large. 
These fluctuations will affect the propagator $G_F(x,y|g)$ and will 
modify it. 
If we again write $g_{ik}$ as $(\bar g_{ik} + h_{ik})$ and average over 
the fluctuations $h_{ik}$, then the effective propagator will be
\begin{equation}
G_F(x,y|\bar g) \equiv \sum\limits_h G_F(x, y |\bar g + h ){\cal P}(h)\label{eqn:haver}
\end{equation}
where ${\cal P}(h)$ is the amplitude for a fluctuation $h_{ik}$, which 
will depend on the `correct' theory of gravity.
We are interested in knowing the modified form of the propagator.

It is, of course, impossible to `derive' the correct propagator which 
takes into account quantum fluctuations of metric. 
To do so, one needs a workable model for quantum gravity which will 
give us ${\cal P}(h)$. Since we do not have this, the  best one can 
do is to take hints from various models for quantum gravity and come 
up with an ansatz. 
This is what I propose to do along the following lines.

The strongest hint is the existence of the length, $L_P \equiv
(G\hbar/c^3)^{1/2}$, which is expected to play a vital role in 
the `ultimate' theory of  quantum gravity. 
Simple thought experiments indicate that it is not possible to devise 
experimental procedures which will measure lengths with an accuracy 
greater than about ${\cal O}(L_P)$~\cite{paddy87}. 
This result suggests that one could think of Planck length as some 
kind of ``zero-point length'' of spacetime. 
In some simple models of quantum gravity, $L_P^2$ does arise as a mean 
square fluctuation to spacetime intervals, due to quantum fluctuations 
of the metric~\cite{paddy85}. 
In more sophisticated approaches, like models 
based on string theory or Ashtekar variables, 
similar results arise in one guise or the other (see 
{\it e.g.},~\cite{rov95,asht92,ashk94,alvrz94,sch96,pol96,dne96}).  
The existence of a fundamental length  implies that processes involving 
energies higher than Planck energies will be suppressed and the 
ultra-violet behavior of the theory will be improved. 
All sensible models for quantum gravity provide some mechanism for good 
ultra-violet behavior, essentially through the existence of a 
fundamental length scale. One direct consequence of such an improved 
behavior will be that the Feynman propagator (in momentum space) will 
acquire a damping factor for  energies larger than Planck energy.

If the ultimate theory of quantum gravity has a fundamental length scale 
built into it, then it seems worthwhile  to use this principle as the 
starting point to obtain a glimpse of the modifications introduced by 
quantum gravity effects at lower energies, provided we can introduce the 
quantum gravity effects through some powerful, general principle. 

With this motivation in mind, let us ask how the  propagation amplitude 
could be modified if there exists a fundamental zero-point length to the
spacetime. 
In equation~(\ref{eqn:equform}), the weightage given for a path of length 
${\cal R}$ is $\exp (-m{\cal R})$ which is a monotonically decreasing 
function of ${\cal R}$. 
The existence of a fundamental length $L_P$ would suggest that paths with 
length ${\cal R} \ll L_P$ should be suppressed in the path integral. 
This can, of course, be done in several different ways by arbitrarily 
modifying the expression in equation~(\ref{eqn:equform}).  
In order to make a specific choice I shall invoke the following 
`principle of duality'. 
I will postulate that the weightage given for a path  should be invariant 
under the transformation ${\cal R}\to L_P^2/{\cal R}$. Since the original 
path integral has the factor $\exp (-m{\cal R})$ we have to introduce the
additional factor $\exp (-mL_P^2/{\cal R})$. 
We, therefore, modify equation~(\ref{eqn:equform}) to
\begin{equation}
G_F(x,y|g)= \sum \exp\left(-m ({\cal R} + {L_P^2 \over {\cal R}}) \right)\label{eqn:stpt}
\end{equation}
I will take this to be the basic postulate arising from the `correct' 
theory  of quantum gravity. 
It may be noted that the `principle of duality' invoked here is
similar to that which arises in string theories (though not
identical)~\cite{ashk94,alvrz94,sch96,pol96,dne96}. 
In fact we may think of equation~(\ref{eqn:stpt}) as a result of performing 
the averaging on the right hand side of equation~(\ref{eqn:haver}). 
Since I do not know ${\cal P}(h)$, this result is a postulate at present. 
It is also the simplest realization of duality for a free particle; 
we have demanded that the existence of a weightage factor $\exp (-ml)$
necessarily requires the existence of another factor $\exp (-mL_P^2/l)$. 
We shall now study the consequences of the modifications we have 
introduced. 

To do this we need to evaluate the path integral in 
equation~(\ref{eqn:stpt}). 
It turns out that this can indeed be done (see section~\ref{sec:dualinv}) 
and the result is quite simple to state:
\begin{equation}
G_F(x,y|g)= \sum \exp\left(-m ({\cal R} 
+ {L_P^2 \over {\cal R}}) \right)
= \int\limits^{\infty}_0 d \tau 
\exp\left(-m^2\tau - {L^2_p \over \tau}\right)
K (x, x', \tau | \bar g)\label{eqn:finres}
\end{equation}
Our modification merely changes the weightage given to a path of proper
time $\tau$ from $\exp(-m^2\tau)$ to $\exp(-m^2\tau-L_P^2/\tau)$ in 
Schwinger's prescription.

This result has an interesting interpretation. 
It is well-known that the Kernel $K(x,y;\tau|g)$ has the 
DeWitt-Schwinger expansion of the form
\begin{equation}
K(x,y;\tau|\bar g) 
=\left( {1 \over 4 \pi \tau }\right)^{D/2} 
\exp\left(-{(x-y)^2 \over 4 \tau}\right)\left[1+.....\right]
\label{eqn:dew1}
\end{equation}
where the dots.....  represent metric-dependent corrections. 
Using equation~(\ref{eqn:dew1})  in equation~(\ref{eqn:finres}) 
we can write our propagator as
\begin{equation}
G_F(x,y|\bar g)
=  \int\limits^{\infty}_0 d \tau e^{-m^2\tau} 
\left( {1 \over 4 \pi \tau }\right)^{D/2} 
\exp \left(-{(x-y)^2 +4 L^2_p\over 4 \tau}\right)[1 +....]
\end{equation}
Thus the net effect of our modification is to add a ``zero-point length"
$4L^2_p$ to $(x-y)^2$ in the original propagator. 
{\it The postulate of duality used in the path integral is identical to 
the postulate of such a zero-point length}. 
This is one of the key results of this paper and---as far as I
can see---this connection is far from obvious.

In the case of flat background spacetime, the terms indicated by the 
dots ...... vanish and the propagator is given by
\begin{equation}
G_F(x)
= \left({1 \over 4\pi }\right)^{D/2}
\int_0^{\infty}{ ds \over s^{D/2}} 
\exp \left( -m^2s - {1 \over 4s} (x^2 + l^2) \right)\label{eqn:gxlp} 
\end{equation}
where we have set $y = 0, \tau = ms$ and defined $l\equiv 2L$. 
To see the effect of our new term, we may Fourier transform this 
expression with respect to $x$ giving:

\begin{equation}
\tilde G({\bf p})= 
\int\limits^{\infty}_0 ds \exp\left(-(p^2 + m^2)s - {l^2 \over 4s}\right)
\label{qflmom}
\end{equation}
When $l=0$, this gives the conventional Feynman propagator in Fourier space
$(p^2+m^2)^{-1}$.  When $l\neq 0$ the integral can be performed to give
\begin{equation}
\tilde G({\bf p})= K_1(l \sqrt{p^2 + m^2}) {l \over \sqrt{p^2 + m^2}}
\end{equation}
where $K_1(z)$ is the modified Bessel function. The limiting forms of this
expression are
\begin{equation}
\hat G({\bf p}) \rightarrow \cases{ (p^2+m^2)^{-1} & (for $l
\sqrt{p^2+m^2} \ll 1$)\cr
\exp(- l \sqrt{p^2+m^2}) & (for $l \sqrt{p^2+m^2} \gg 1$).\cr}
\end{equation}
which clearly shows the suppression of energies higher than Planck energies.

The rest of the paper is organized as follows. 
In section~\ref{sec:warmup}, I illustrate how the path integral 
can be rigorously defined using a D-dimensional lattice and 
limiting procedure. 
This `warm-up' exercise shows how the standard result~(\ref{eqn:equform}) 
arises and sets the stage for the main analysis of the paper. 
In section~\ref{sec:dualinv}, I evaluate the modified path integral using 
the same technique and obtain equation~(\ref{eqn:finres}). 
Some of the implications are discussed in section~\ref{sec:cnclsns}.

\section{Warm up: Feynman propagator from  sum over paths}\label{sec:warmup}
\subsection{Rigorous evaluation of path integral}\label{subsec:rigor}

In defining the path integral in nonrelativistic quantum mechanics, 
we discretize the time axis, define the path integral with a non-zero 
spacing $\epsilon$ and finally take the limit of $\epsilon$ going to 
zero. 
To define the path integral in D-dimensions we can use a similar 
procedure. 
We will work in the Euclidean space and introduce a cubic lattice with 
spacing $\epsilon$. 
The path integral will be defined on the lattice and then we will take 
the limit of $\epsilon \rightarrow 0$. To obtain a finite value in the 
limit of $ \epsilon \rightarrow 0$ we have to choose the measure and 
the mass parameter, $m$,  which varies  in a specific fashion with 
$\epsilon$. 
This can be done fairly easily and the final expression will agree with 
the standard Feynman propagator for a free scalar field. 
The calculation proceeds as follows:

We will work directly in the Euclidean space  of D-dimensions. 
In this section we are primarily interested in the issues of 
principle, regarding the measure for the path integral, and 
will consider the path integral for a free particle. 
We have to, therefore, evaluate
\begin{equation}{\cal G}_E({\bf x_2, x_1};\mu_0)=
\sum_{all\,{\bf x}(t)}\exp
-m\,l[{\bf x}(t)] \label{eqn:lat}
\end{equation}
in the Euclidean sector, where $l$ is 
\begin{equation}
l({\bf x_2,x_1})
=\int^1_0ds\left |\left({d{\bf x}\over ds}\right)^2\right |^{1/2}
\end{equation}
is just the length of the curve ${\bf x}(s)$, connecting
${\bf x}(0)={\bf x}_1$ and ${\bf x}(1)={\bf x}_2$.

This quantity can be defined through the following limiting 
procedure:~Consider a lattice of points in a D-dimensional 
lattice with a uniform lattice spacing of $\epsilon$. 
We will work out ${\cal G}_E$ in the lattice and will then take 
the limit of $\epsilon\to 0$ with suitable measure $M(\epsilon)$. 
To obtain a finite result, it is also necessary to  treat $m$ 
(which is the only parameter in the problem) as  a function 
$\mu(\epsilon )$ of lattice spacing where the functional form 
is to be chosen in such a way to ensure finiteness of the 
continuum result. 
We will reserve the symbol $m$, for the  value of this function 
in the continuum limit. 
Thus we  will define the path integral result as a limit:
\begin{equation}
{\cal G}({\bf x_2,x_1};m)=
\lim_{\epsilon\to 0}
\left[M(\epsilon){\cal G}({\bf x_2, x_1};
\mu(\epsilon))\right]\label{eqn:pathintres}
\end{equation}
where  the functions $M(\epsilon)$ and $\mu(\epsilon)$ are to be
chosen so as to ensure finiteness. 
The rationale for this expression arises from the following point 
of view:~We treat the continuum space as a limit of a lattice with 
the lattice spacing $\epsilon$ going to zero. We now construct a 
sequence of path integrals parametrized by the spacing $\epsilon$ 
by choosing certain functions $\mu(\epsilon)$ and $M(\epsilon)$ and 
{\it define}\/ the continuum path integral as the limit of this 
sequence. 
We shall show later that this limit exists only if $\mu(\epsilon) 
\simeq (ln 2D) /\epsilon$ and $M(\epsilon) \simeq (2D)^{-1} 
\epsilon^{-(D-2)}$ near $\epsilon \simeq 0$. 
The form of   $\mu(\epsilon), M(\epsilon)$ for $\epsilon$ far away 
from zero, of course, makes no difference to the result we are after.

In a lattice with spacing of $\epsilon$, (\ref{eqn:lat}) can be evaluated 
in a straightforward manner.
Because of the translation invariance of the problem, ${\cal G}_E$
can only depend on ${\bf x}_2-{\bf x}_1$; so we can set ${\bf x}_1=0$
and call ${\bf x}_2=\epsilon {\bf R}$ where ${\bf R}$ is a $D$-dimensional 
vector with integral components: ${\bf R}=(n_1,n_2,n_3\cdots n_D)$. 
Let $C(N,{\bf R})$ be the number of paths of length $N\epsilon$ 
connecting the origin to the lattice point $\epsilon{\bf R}$.  
Since all the paths contribute a term [$\exp-\mu(\epsilon)(N\epsilon)$] 
to (\ref{eqn:pathintres}), we get, 
\begin{equation}
{\cal G}_E({\bf R};\epsilon)=
\sum^{\infty}_{N=0}C(N;{\bf R})\exp~[-\mu(\epsilon)N\epsilon]
\end{equation}
The generating function determining 
$C(N;{\bf R})=C(N;n_1, n_2,.....n_D)$ can be calculated easily by 
the following arguments:~Consider any particular path connecting 
the origin to the lattice point ${\bf R}$. Suppose that this path 
takes $r_1$ steps towards positive direction (`right') in first 
axis and $l_1$ steps towards negative direction (`left') in the 
first axis. 
Then $n_1 = r_1 -l_1$; similarly $n_i=r_i - l_i$. 
The number of paths with specified number of $(r_i, l_i)$ for 
$i = 1, ....D$ is just the number of ways of choosing the integers 
$(r_1,...r_D, l_1,...l_D)$ with $\sum r_i + \sum l_i = N$. 
This is given by the coefficient of the polynomial expansion
\begin{equation}
\left( x_1 + x_2 ....x_D + y_1 + y_2 + .... y_D \right)^N 
= \sum Q(N;r_i, l_i)
x_1^{r_1}....x_D^{r_D} y_1^{l_1} ...y_D^{l_D}\label{eqn:expnsn} 
\end{equation} 
In our problem, we allow $(r_i, l_i)$ also to vary keeping 
$r_i-l_i = n_i$ fixed for each $i$. The number of paths with 
this property is clearly given by using the above expression 
with $y_i = (1/x_i)$. 
Then we get
\begin{equation}
\left( x_1 + x_2 ...x_D + {1 \over x_1} 
+ {1 \over x_2}...{1 \over x_D} \right)^N 
= \sum C(N; n_1, n_2,...n_D)x_1^{n_1}...x_D^{n_D} 
\end{equation} 
The expansion of the left hand side gives the generating function 
for $C(N;{\bf R})$.
For further manipulation, it is convenient to set $x_1 = e^{ik_1}, x_2 
= e^{ik_2}, ....x_D= e^{ik_D}$. 
Then we can write
\begin{equation}
F^N\equiv\left[e^{ik_1}+e^{ik_2}+\cdots e^{ik_D}+
e^{-ik_1}+\cdots e^{-ik_D}\right]^N
=\sum_{{\bf R}} C(N;{\bf R})e^{i{\bf k.R}}
\end{equation}
Therefore,
\begin{eqnarray}\sum_{\bf R}e^{i{\bf k.R}}
{\cal G}_E({\bf R};\epsilon)
&=&\sum^{\infty}_{N=0}\sum_{{\bf R}}C(N;{\bf R})
e^{i{\bf k.R}}\exp~[-\mu(\epsilon)N\epsilon]\nonumber\\
&=&\sum^{\infty}_{N=0}e^{-\mu(\epsilon)\epsilon  N}
F^N=\sum^{\infty}_{N=0}\left[Fe^{-\mu(\epsilon)\epsilon}
\right]^N\nonumber\\
&=&\left[1-Fe^{-\mu(\epsilon)\epsilon}\right]^{-1}\label{eqn:Ceval}
\end{eqnarray}
Inverting the Fourier transform, we have
\begin{equation} {\cal G}_E({\bf R};\epsilon)=\int
{d^D{\bf k}\over (2\pi)^D}
{e^{-i{\bf k.R}}\over (1-e^{-\mu(\epsilon)\epsilon}F)}
=\int{d^D{\bf k}\over (2\pi)^D}
{e^{-i{\bf k.R}}\over (1-2e^{-\mu(\epsilon)\epsilon}
\sum^D_{j=1}\cos k_j)}\label{eqn:FTinv}
\end{equation}
Converting to the physical length scales ${\bf x}=\epsilon{\bf R}$ and
${\bf p}=\epsilon^{-1}{\bf k}$ we get
\begin{equation}
{\cal G}_E({\bf x};\epsilon)=
\int{\epsilon^D d^D{\bf p}\over(2\pi)^D}
{e^{-i{\bf p.x}}\over (1-2e^{-\mu(\epsilon)\epsilon}
\sum^D_{j=1}\cos p_j \epsilon)}
\end{equation}
We are now ready to take the limit of zero lattice spacing. 
As $\epsilon\to 0$, the denominator of the integrand
becomes
\begin{equation}
1-2e^{-\epsilon\mu(\epsilon)}
(D-{1\over 2}\epsilon^2|{\bf p}|^2)
=1-2De^{-\epsilon\mu(\epsilon)}
+\epsilon^2 e^{-\epsilon\mu(\epsilon)}|{\bf p}|^2
= \epsilon^2 e^{-\epsilon\mu(\epsilon)}
\left[|{\bf p}|^2+{1-2De^{-\epsilon\mu(\epsilon)}\over 
\epsilon^2 e^{-\epsilon\mu(\epsilon)}}\right]
\end{equation}
so that we will get, for small $\epsilon$,
\begin{equation}
{\cal G}_E({\bf x};\epsilon)
\simeq \int{d^D{\bf p}\over (2\pi)^D}
{A(\epsilon)e^{-i{\bf p.x}}\over |{\bf p}|^2+B(\epsilon)}
\end{equation}
where
\begin{equation}
 A(\epsilon)= \epsilon^{D-2}e^{\epsilon\mu(\epsilon)}; 
\quad B(\epsilon)={1\over \epsilon^2}
\left[e^{\epsilon\mu(\epsilon)}-2D\right]
\end{equation}
The continuum theory has to be defined in the limit of $\epsilon\to 0$ 
with some measure $M(\epsilon)$; that is we want to obtain
\begin{equation}
{\cal G}_E({\bf x};m)|_{{\rm continuum}}
=\lim_{\epsilon\to 0} \left\{M(\epsilon)
{\cal G}_E({\bf x};\epsilon)\right\}
\end{equation}
The choice of the measure is dictated by the requirement that the right
hand side should be finite in this limit. 
We now demand
\begin{equation}
\lim_{\epsilon\to 0} \left[{1\over \epsilon^2}
\left(e^{\epsilon\mu(\epsilon)}-2D\right)\right]=m^2\label{eqn:condtn}
\end{equation}
and
\begin{equation}
\lim_{\epsilon\to 0}
\left[M(\epsilon)\epsilon^{D-2}e^{\epsilon\mu(\epsilon)}\right]
=1\label{eqn:contwo}
\end{equation}
The first condition implies that, near $\epsilon\approx 0$,
\begin{equation}
\mu(\epsilon)\approx {\ln 2D\over \epsilon}
+{m^2\over 2D}\epsilon
\approx {\ln 2D\over \epsilon}\label{eqn:latmass}
\end{equation}
Using this in the second condition~(\ref{eqn:contwo}), we can determine 
the measure as
\begin{equation}
M(\epsilon)={1\over 2D}{1\over \epsilon^{D-2}}\label{eqn:meas}
\end{equation}
With this choice, we get
\begin{equation}
G_E({\bf x}; m)\equiv 
\lim_{\epsilon\to 0}{\cal G}_E
({\bf x};\epsilon)M(\epsilon)=\int
{d^D{\bf p}\over(2\pi)^D}
{e^{-i{\bf p.x}}\over |{\bf p}|^2+m^2}
\end{equation}
which is the  standard Feynman propagator.
This analysis gives a rigorous meaning to the nonquadratic path 
integral with a square root and also illustrates the role played 
by the choice of the measure. 
In the continuum limit, we have only one length scale $m^{-1}$; 
this fact suggests that the right hand side of~(\ref{eqn:condtn}) 
should scale as $m^2$. 
Setting the proportionality constant to unity should be thought 
of a (part of) choice of measure. 
Similarly, $M(\epsilon)$ can be multiplied by any finite quantity. 
The choice in~(\ref{eqn:contwo}) should also be considered as part of 
the definition of measure.

To connect this expression with Schwinger's proper time 
representation is easy.
By writing $(|{\bf p|}^2 + m^2)^{-1}$ as
\begin{equation}
{1 \over  |{\bf p|}^2 + m^2} 
= \int\limits^{\infty}_0 d\tau\; 
e^{- \tau(m^2 + |{\bf p|}^2)} \label{eqn:exid}
\end{equation}
and doing the integral over ${\bf p}$, we get 
\begin{equation}
G_E ({\bf x}; m ) 
= \sum \exp (- m{\cal R}) 
= \int_0^{\infty} {d \tau \over (4 \pi \tau)^{D/2}} 
e^{-m^2 \tau} \exp (- {|x^2| \over 4 \tau} )\label{eqn:gint}
\end{equation}
Part of the integrand can be expressed as an ordinary quadratic 
path integral:
\begin{eqnarray}
K({\bf x}, {\bf y}; \tau)
&\equiv & \int {\cal D} x \exp \left(- {1 \over 4} 
\int\limits^{\tau}_0 \dot x^i \dot x_i ds \right)\nonumber\\
&=&\left( {1 \over 4 \pi \tau} \right)^{D/2} 
\exp \left( - {({\bf x} - {\bf y})^2 \over 4 \tau} \right) 
\end{eqnarray}
where we have shifted the origin to ${\bf y}$. 
Using this in~(\ref{eqn:gint}), we get the final result, 
quoted in~(\ref{eqn:Ceval}):
\begin{equation}
\sum \exp [-m{\cal R}(x,y)] 
= \int\limits^{\infty}_0 d \tau e^{-m^2 \tau} 
\int {\cal D} x \exp \left( - {1 \over 4} 
\int\limits^{\tau}_0 \dot x_i \dot x^i ds \right)
\end{equation}

\subsection{Physical interpretation}\label{subsec:physint}

The above analysis relates a nonquadratic path integral (containing 
a square root) to a standard quadratic path integral. 
This result has a simple physical interpretation, which is worth 
emphasizing. 
Consider the standard path integral Kernel $K({\bf x}, {\bf y}, \tau)$ 
in quantum mechanics, defined through the Hamiltonian form of the action:
\begin{equation}
K({\bf x}, {\bf y}, t)=\sum_{{\bf x}(t)}
\sum_{{\bf p}(t)}\exp {i\over \hbar}\int
dt({\bf p.\dot x}-H({\bf p,x}))\label{eqn:jacobipi}
\end{equation}
with $H\geq 0$.
From the principles of quantum mechanics, we would expect the Fourier 
transform
\begin{equation}
B({\bf x}, {\bf y}; E) 
\equiv \int_0^{\infty} K ({\bf x}, {\bf y}, ; E ) 
e^{iEt} dt \label{eqn:debk}
\end{equation}
to give the amplitude for the particle to propagate from ${\bf y}$ to 
${\bf x}$ with energy $E$. (Only $t \geq 0$ is relevant in the Fourier 
transform (\ref {eqn:debk}), since $K$ is taken to vanish for $t < 0$). 
But the trajectory of a classical particle with fixed energy can be 
described using the Jacobi action:
\begin{equation}
A_{\rm Jacobi} = \int\limits^t_0 dt' \sqrt{2m_0(E-V)|\dot {\bf x}|^2}
\end{equation}
We will therefore expect the relation
\begin{eqnarray}
\sum\limits_{\rm paths} \exp i \int\limits^t_0 dt' 
\sqrt{2m_0(E-V)|\dot {\bf x}|^2}&=&B({\bf x}, {\bf y}; E)\nonumber\\
&=& \int\limits^{\infty}_0 dt \quad  e^{iEt} 
\int {\cal D} {\bf x} \exp i \int_0^t ({1 \over 2} m_0 \dot x^2 -V)dt' \label{eqn:myre}
\end{eqnarray}
to hold, thereby expressing a square root path integral in terms of a 
quadratic path integral. 
Taking $m_0 = (m/2), V=0, E= m$ and switching to Euclidean sector gives 
the result
\begin{equation}
\sum\limits_{\rm paths} \exp [- m{\cal R} ({\bf x}, {\bf y} )] 
= \int_0^{\infty} dt \quad e^{imt} \int {\cal D} x 
\exp \left[ - {m \over 4} \int\limits^t_0 \dot x^2 dt' \right] 
\end{equation}
which is the same as (2) after the rescaling $t =m\tau$ and continuing to 
the Euclidean sector. The choice of $E = mc^2$ shows that the energy of 
the particle in the rest frame is on the mass shell.

To prove the result~(\ref{eqn:myre}), we need the following path integral identities:
\begin{equation}
\delta(f(t))=\sum_{\lambda(t)}\exp i\int dt\lambda(t)f(t)\label{eqn:one}
\end{equation}
\begin{equation}\sum_{{\bf p}}\exp i\int dt
\left[{\bf p}.{\bf\dot x}+a(t)p^2\right]=\exp i\int dt{\dot x^2\over 4a(t)}\label{eqn:two}
\end{equation}
\begin{equation}
\sum_{\lambda(t)}\exp i\int dt
\left(\lambda(t)a(t)+{b(t)\over \lambda(t)}\right)=
\exp i\int dt\left[-4ab\right]^{1/2}\label{eqn:three}
\end{equation}
\n 
The first result is merely the definition of delta functional;~the 
second and third can be obtained in the Euclidean Sector by standard 
time slicing techniques and can be analytically continued. 
They are direct generalizations of the corresponding results of ordinary
integrals. 

Introducing into the integrand of~(\ref{eqn:jacobipi}), the 
`expansion of unity' in the form:
\begin{equation}
1=\int^{\infty}_0 dE\delta(E-H({\bf p,x}))
\end{equation}
we get:
\begin{eqnarray}
 K({\bf x}, {\bf y};t)&=&
\int_0^{\infty} dE\sum_{{\bf x}}\sum_{{\bf p}}\delta
(E-H({\bf p,x}))\exp {i\over \hbar}
\int dt ({\bf p.\dot x}-H({\bf p,x}))\nonumber\\
&=& \int_0^{\infty} dE \sum_{{\bf x}}\sum_{{\bf p}}\delta
(E-H)e^{-iEt}\exp i\int dt ({\bf p.\dot x})
\end{eqnarray}
So
\begin{equation}
\int^{\infty}_0K({\bf x}, {\bf y}; t) 
e^{iEt} dt\equiv B({\bf x}, {\bf y}; E)
=\sum_{{\bf x}}\sum_{{\bf p}}\delta
\left({{\bf p}^2\over 2m}+ V({\bf x})-E\right)
\exp i\int dt{\bf p.\dot x}
\end{equation}
\n We now express the delta functional using equation~(\ref{eqn:gint}):
\begin{equation}\delta
\left({{\bf p}^2\over 2m}+V({\bf x})-E\right)=
\sum_{\lambda(t)}\exp i\int
\lambda(t)
\left[{p^2\over 2m}+V({\bf x})-E\right]dt.
\end{equation}
Then
\begin{eqnarray}
B({\bf x}, {\bf y}; E)&=&
\sum_{{\bf x}}\sum_{\lambda(t)}\exp i \int dt\lambda(t)
\left[V({\bf x})-E\right]
\sum_{{\bf p}}\exp i\int dt
\left[{\bf p.\dot x}+{\lambda (t)\over 2m} p^2\right]\nonumber\\
&=& \sum_{{\bf x}}\sum_{\lambda(t)}\exp i\int dt
\lambda(t)\left[V({\bf x})-E\right].\exp
i\int dt{1\over 2} {m\over \lambda(t)}{\bf \dot x}^2 \nonumber\\
&=& \sum_{{\bf x}}\sum_{\lambda(t)}\exp i\int dt
\left[\lambda(t)(V({\bf x})-E)+{m\over 2\lambda(t)}
{\bf \dot x}^2\right] \nonumber\\
&=& \sum_{{\bf x}}\exp i\int dt
\sqrt{2m(E-V)|{\bf \dot x}|^2}.
\end{eqnarray}
In arriving at the second equality, we have used~(\ref{eqn:two}) and 
in arriving at the last equality we have used~(\ref{eqn:three}). 
This proves the result quoted above.

To summarize, we have demonstrated how path integrals involving square 
roots can be given a rigorous definition---using a lattice regularization 
scheme---in section~\ref{subsec:rigor}. 
This definition of path integral is given a more intuitive interpretation 
in section~\ref{subsec:physint}. 
We shall now work out the modified path integral along the same lines.

\section{Feynman propagator with duality invariant path 
integral}\label{sec:dualinv}

We shall now turn to the main task of the paper, viz. evaluation of the 
modified path integral in~(\ref{eqn:stpt}). 
It is easy to see that the lattice version now becomes 
\begin{equation} 
{\cal G} ({\bf R}, \epsilon) 
= \sum_{N=0}^\infty C(N, {\bf R}) 
\exp \left[ - \mu (\epsilon) \epsilon N 
- {\lambda (\epsilon) \over \epsilon N} \right],\label{eqn:latmod}
\end{equation} 
where $\lambda(\epsilon)$ is a lattice parameter which will play the role 
of $(m L_P^2)$ in the continuum limit. 
This replaces equation~(\ref{eqn:expnsn}) of previous analysis. 
After evaluating $G({\bf R}, \epsilon)$, we multiply it by a measure 
${\cal M }(\epsilon )$ and take the limit $\epsilon \rightarrow 0$. 
The functions ${\cal M} (\epsilon)$, $\mu(\epsilon)$  $\lambda (\epsilon)$ 
are to be chosen so that, in the continuum limit $\mu$ corresponds to the 
mass $m$ and $\lambda$ to $(mL^2_p)$. 
Since we expect the result to have the correct limit as 
$L_p \rightarrow 0$, we anticipate that  take the form 
of $\mu(\epsilon)$ will be as given by~(\ref{eqn:meas}).

To evaluate this path integral on the lattice we again begin 
with  the generating function for $C(N,{\bf R})$, given by
equation~(\ref{eqn:Ceval})
\begin{equation}
F^N  \equiv  \sum_\R C(N; \R) e^{i{\bf k}\cdot \R}
= \biggl(e^{ik_1} + e^{ik_2} + \cdots 
+ e^{ik_D}
+\, e^{-ik_1} + e^{-ik_2} + \cdots + e^{-ik_D}\biggl)^N.
\end{equation}
This now leads to the expression: 
\begin{eqnarray}
\sum_\R e^{i{\bf k}\cdot \R} {\cal G}(\R, \epsilon) 
&=& \sum_{N=0}^\infty e^{-\mu\epsilon N - (\lambda / \epsilon N)} 
\sum_\R C(N,\R) e^{i{\bf k}\cdot \R}\nonumber\\ 
&=& \sum_{N=0}^\infty e^{-N (\mu \epsilon - \ln F) 
- (\lambda/\epsilon N)}.\label{eqn:qsumone}
\end{eqnarray}
Thus, our problem reduces to evaluating the sum of the form 
\begin{eqnarray}
S(a,b) &\equiv& \sum_{n=0}^{\infty} 
\exp\left( -a^2 n - {b^2\over n}\right)\nonumber\\
&=& \sum_{n=1}^{\infty} \exp\left(-a^2 n 
- {b^2\over n}\right).\label{eqn:qamb}
\end{eqnarray}
which is more complicated than the geometric progression of 
equation~(\ref{eqn:FTinv}). 
Fortunately this expression can be evaluated by some algebraic tricks 
(see appendix~\ref{app:sumeval}) 
and the answer is 
\begin{eqnarray}
S(a,b) &=&\int_{0}^{\infty} {q dq\over 2b^2} {J_0(q) 
e^{-(a^2 + q^2/4b^2)}\over \left[1 
- e^{-(a^2 + q^2/4b^2)}\right]^2}\nonumber\\ 
&=& {1\over \left( 1-e^{-a^2}\right)} 
- \int_0^\infty dq {J_1(q) \over 
\left[1 - e^{-(a^2 + q^2/4b^2)}\right]},\label{eqn:qanss}
\end{eqnarray}
where $J_\nu (x)$ is the Bessel function of order $\nu$. 
The first form of the integral  shows that the expression is well 
defined while the second form has the advantage of separating out 
the $b$-independent part as the first term. 
(Note that the two summations in~(\ref{eqn:qamb}) will differ by 
unity if $b=0$; the results in~(\ref{eqn:qanss}) will go over to 
the second summation in~(\ref{eqn:qamb}) if the limit $b\to 0$ is 
taken.)
In our case, $b^2=(\lambda/\epsilon)$ and $a^2 = \mu \epsilon - 
\ln F:$ so we get:
\begin{equation}
S(a,b) 
=\int_0^{\infty} q dq 
J_0 (q)\left\lbrace {\epsilon \over 2 \lambda} 
{F \exp(-\mu \epsilon- {q^2 \epsilon \over 4 \lambda}) 
\over \left[ 1 - F \exp(-\mu \epsilon 
- {q^2 \epsilon \over 4 \lambda})\right]^2}\right\rbrace
\end{equation}
This gives
\begin{equation}
G({\bf R}) 
= \int {d^Dk\over (2\pi)D} \int_0^{\infty} q dq 
J_0 (q) e^{-i{\bf k . R}}\left\lbrace {\epsilon \over 2 \lambda} 
{F \exp (-\mu \epsilon -  {q^2 \epsilon \over 4 \lambda})  
\over \left[ 1 - F \exp(-\mu \epsilon 
-{q^2 \epsilon \over 4\lambda})\right]^2} \right\rbrace
\end{equation}
Rescaling back to $ {\bf x} = \epsilon R, {\bf p}
= \epsilon^{-1} {\bf k}$, we find
\begin{equation}
G({\bf x}) 
= \int {\epsilon^D d^Dp \over (2 \pi )^D} 
\int_0^{\infty} q dq J_0 (q)
\left\lbrace {\epsilon \over 2 \lambda} 
{H e^{-i{\bf p}. {\bf x}} 
\over (1 -H)^2 }\right\rbrace\label{eqn:gxdualinv}
\end{equation}
with
\begin{equation}
H \equiv F \exp (- \alpha \epsilon) 
\equiv F \exp \left \lbrace - \mu \epsilon - {q^2 \over 4} 
{\epsilon \over \lambda } \right \rbrace ; \quad \alpha \epsilon
= \mu (\epsilon ) \epsilon + {q^2 \over 4} {\epsilon \over \lambda}
\end{equation}
This expression is dimensionless; we now take the $\epsilon 
\rightarrow 0$ limit, to get
\begin{eqnarray}
1-H
&=& 1-2 e^{-\epsilon \alpha } \sum\limits^D_{i = 1} \cos \epsilon p_{i} 
\approx 1-2 e^{- \epsilon \alpha} \left\lbrack D 
- {1 \over 2} \epsilon^2 p^2 \right \rbrack\nonumber\\ 
&=& \epsilon^2 e^{-\alpha \epsilon} [p^2 
+ {e^{\alpha \epsilon} \over \epsilon^2} (1 -2 De^{-\alpha \epsilon})] 
\end{eqnarray}
So we can write, retaining leading terms:
\begin{equation}
G({\bf x})
\cong  \int {\epsilon^D d^D p \over (2 \pi )^D} 
\int_0^{\infty} q dq J_0 (q) e^{-i{\bf p} . {\bf x}}
\left\lbrack {e^{\alpha \epsilon} \over 2 \lambda \epsilon^3} 
{2D \over (p^2 + B)^2}\right\rbrack 
\end{equation}
where $B(\epsilon)$ is defined as 
\begin{equation}
B = {1 \over \epsilon^2 } 
\left\lbrace e^{\mu \epsilon 
+ {q^2 \over 4} {\epsilon \over \lambda}} - 2D \right\rbrace
\end{equation}
Consider now the $\epsilon \rightarrow 0$ limit of $B$; using (\ref{eqn:latmass}), we have $e^{\mu \epsilon} \approx 2D 
+ m^2 \epsilon^2$ for small $\epsilon$. 
So  
\begin{equation}
B \approx {1 \over \epsilon^2} 
\left\lbrace 2D (e^{{q^2 \over 4}{\epsilon \over \lambda}} - 1 ) 
+m^2 \epsilon^2 e^{{q^2 \over 4}{\epsilon \over \lambda}}\right\rbrace
\end{equation}
For the first term to be finite at $\epsilon \rightarrow 0 $ limit, 
we need the small - $\epsilon$ dependence to be of the form:
\begin{equation}
\exp\left( {q^2 \epsilon \over 4 \lambda} \right)-1 
\approx  A_1 (q) \epsilon^2
\end{equation}
This implies that
\begin{equation}
\exp\left( {q^2 \epsilon \over 4 \lambda} \right) \approx 1 
+ A_1 \epsilon^2 \approx 1 + {q^2 \over 4 } {\epsilon \over \lambda}
\end{equation}
giving
\begin{eqnarray}
B &\approx & {1 \over \epsilon^2} 
\left\lbrace 2 DA_1 \epsilon^2 + m^2 \epsilon^2 
(1 + {\cal O}(\epsilon^2)....) \right\rbrace\nonumber\\
&=&(2DA_1 + m^2)
\end{eqnarray}
Further, since $A_1(q)\epsilon^2 = (q^2/4) (\epsilon / \lambda)$, we 
need $\lambda$ to scale as $\lambda \approx (q^2 /4A_1)(1/\epsilon)$
 as $\epsilon \rightarrow 0$. Since $\lambda(\epsilon)$ is to be 
independent of $q$, we must have $A_1 (q) = (L^{-2}q^2/2D)$ with 
$L = {\rm constant}$.  
Then $\lambda (\epsilon) \approx (L^2/4)(1/\epsilon)(2D)$ as $\epsilon
\rightarrow 0$. We thus find that, near $\epsilon \cong 0$, we need 
\begin{equation}
B \cong m^2 + {q^2 \over L^2} ; \quad e^{\alpha \epsilon} 
= e^{\mu \epsilon} e^{q^2 \epsilon \over 4 \lambda} 
\cong 2D; \quad \lambda \epsilon^3 \approx {L^2 \over 4} \epsilon^2
\end{equation}
Putting everything into (\ref{eqn:gxdualinv}), we get
\begin{eqnarray}
G({\bf x})
&=& \int {\epsilon^D d^Dp \over (2 \pi )^D}
e^{-i{\bf p} . {\bf x}} 
\int_0^{\infty} {q dq J_0 (q) \over L^2 \epsilon^2}
{2 (2D)^2 \over (p^2 + m^2 + q^2 / L^2)^2 }\nonumber\\
&=& 4 \epsilon^{D - 2} D^2  
\int {d^Dp \over (2 \pi)^D}e^{-i{\bf p} . {\bf x}}
\int_0^{\infty} {2qJ_0(q)L^2dq \over [L^2(p^2 + m^2) + q^2 ]^2}
\end{eqnarray}
We now  choose the measure $M(\epsilon)$ such that
\begin{equation}
\lim_{\epsilon \to 0 }4 \epsilon^{D-2} D^2M(\epsilon) = 1
\end{equation}
Then we get the final result:
\begin{equation}
G({\bf x})
= \int {d^Dp \over (2 \pi)^D} e^{-i{\bf p} . {\bf x}}
\int_0^{\infty} { 2qJ_0(q)L^2dq \over [L^2(p^2 + m^2) + q^2 ]^2}
\end{equation}

We have thus successfully defined the path integral in~(\ref{eqn:stpt}) 
using a lattice regularization procedure. 
Note that we now needed 3 functions $M(\epsilon), \mu (\epsilon)$ 
and $\lambda(\epsilon)$. 
Of these, $M(\epsilon)$ and $\mu(\epsilon)$ was required even in the 
standard free particle case, discussed in section~\ref{sec:warmup}. 
In fact, we are using the same functional form $\left(M(\epsilon) 
\propto \epsilon^{2-D}, \mu (\epsilon) \propto \epsilon^{-1}\right)$ 
for these functions near $\epsilon \approx 0$. 
The new entity needed now is $\lambda(\epsilon)$ which should 
correspond to $(mL^2_p)$ in the continuum limit. 
This function scales as $\lambda(\epsilon) \propto (L^2/\epsilon)$ 
near $\epsilon \approx 0$. 
At this stage we can only say that $L \propto L_p$; the proportionality 
constant, as usual, cannot be determined by considerations of measure; 
we shall say more about this later.

Our result can be recast in more useful forms. 
To begin with, the momentum space propagator is given by
\begin{equation}
G({\bf p}) 
\equiv \int d^D {\bf x}  G({\bf x})e^{i{\bf p}.{\bf x}} 
= \int\limits^{\infty}_0 {2qJ_0(q)L^2dq \over [q^2 + L^2 (p^2 + m^2)]^2}
\end{equation}
Using the identity
\begin{equation}
\int_0^{\infty} dz {zJ_0(z) \over (z^2 + Q^2)^2} 
= {K_1(Q) \over 2Q}
\end{equation}
where $K_1(Q)$ is the modified Bessel function, we get:
\begin{equation}
G({\bf p})
= {1 \over \sqrt {p^2 + m^2}} K_1 (L \sqrt {p^2 + m^2}) 
= \cases{(p^2 + m^2)^{-1} &(as $L \rightarrow 0$)\cr
e^{-L \sqrt {p^2 + m^2}} & (as $L \rightarrow \infty$)\cr}
\end{equation}

Clearly, the propagator reduces to standard form $(p^2 + m^2)^{-1}$ 
obtained earlier, when $L^2(p^2 + m^2)\rightarrow 0$. 
By setting $q = L \lambda$ we get
\begin{equation}
G({\bf p}) 
= 2 \int_0^{\infty} {\lambda J_0 (L \lambda) d \lambda
 \over [\lambda^2 + p^2 + m^2 ]^2} 
= - {\partial \over \partial m^2} 
\int_0^{\infty} {2 \lambda J_0 (L \lambda ) d \lambda 
\over [\lambda^2 + p^2 + m^2 ]} 
\end{equation}
Expressing the denominator using the identity~(\ref{eqn:exid}), 
and differentiating with respect to $m^2$, it is easy to show that
\begin{equation}
G(p) = \int_0^{\infty} 2 \tau d \tau e^{- \tau(p^2 + m^2)} W(L, \tau),
\end{equation}
where
\begin{equation}
W(L, \tau)
= \int\limits^{\infty}_0 \lambda d \lambda 
J_0 (L \lambda) e^{- \tau \lambda^2} 
= {1 \over 2 \tau} \exp (-{L^2 \over 2 \tau})
\end{equation}
with the last equality following from a standard identity related to 
Bessel functions. Using this, we can write, 
\begin{equation}
G(p) = \int\limits^{\infty}_0 d \tau 
\exp [- \tau (p^2 + m^2) - {L^2 \over 4 \tau}]
\end{equation}
which has the same form as (10). 
Fourier transforming with respect to ${\bf p}$, we get the key result:
\begin{eqnarray}
G({\bf x})
&=&\int\limits^{\infty}_0 d \tau 
\exp [-\tau m^2 - {x^2 + L^2 \over 4 \tau}]\nonumber\\
&=& \int\limits^{\infty}_0 d\tau 
\exp (-\tau m^2 - {L^2\over 4 \tau}) K(x; \tau).\label{eqn:lastres}
\end{eqnarray}
This is the result quoted in equations~(\ref{eqn:gxlp}) 
and~(\ref{eqn:finres}), if we identify $L^2 = 4 L^2_p$. 
Our definition of limiting procedure only shows that $L \propto L_p$. 
The actual proportionality constant depends on the definition of 
measure and we shall see in the next section why $L^2= 4L^2_p$ is natural.

\section{Generalization to curved spacetime}\label{sec:gencst}

The rigorous way of evaluating equation~(\ref{eqn:stpt}), viz.  to define 
the path integral on a lattice and use a limiting procedure, 
This was done in above  for flat background spacetime. 
It is possible that this procedure can be generalized to curved spacetime. Unfortunately, this procedure hides the extreme simplicity of the result 
in equation~(\ref{eqn:lastres}) and does not make transparent the origin 
of several intermediate results. 
Here, I shall follow a different and simpler route and rederive the result. 
This rederivation suggests a generalization to curved spacetime.

The key idea is that the new factor in the path integral, 
$\exp(-a^2/\R)$, can be expressed in terms of a factor like
$\exp(-b^2\R)$ by performing a gaussian integral. 
The latter factor, of course, can be evaluated in the path integral. 
The gaussian integration will also  produce a $\R^{1/2}$ factor in 
front which needs to be taken care of by doing a two dimensional 
gaussian integration and a differentiation. 
With such elementary algebraic tricks, one can prove 
equation~({\ref{eqn:lastres}).

We start with a slight generalization of equation~(\ref{eqn:equform}),
\begin{equation}
\sum e^{-(m+\alpha){\cal R}}
=\int\limits^{\infty}_0 d \tau K (x', x; \tau |\bar g)  
\exp ({-m(m+\alpha) \tau}) \label{eqn:genform}
\end{equation}
This can be easily proved by lattice techniques used 
in section~\ref{sec:warmup}. 
(See equation~(\ref{eqn:contwo}); redefining the right hand side to be $m(m+\alpha)$ will lead to~(\ref{eqn:genform}).) 
More precisely, this equation {\it defines}\/ the measure used in the 
left hand side of~(\ref{eqn:genform}). 
(This definition is nonstandard in the sense that we have replaced $m$ 
by $(m + \alpha)$ in functional on the left hand side but changed $m^2$ 
to $m(m + \alpha)$ on the right hand side. 
But it is a perfectly valid definition for the measure. 
In fact we can define the right hand side of~(\ref{eqn:contwo}) to be 
in general of the form $m^2F(\alpha/m)$ where $F$ is an arbitrary, 
dimensionless function. 
This is possible because we now have {\it two}\/ dimensional constants $m$ 
and $\alpha$.)  
We now introduce a two real variables $(k_1,k_2)$ with 
$k^2\equiv k_1^2+k_2^2$ and set $\alpha = k^2/m$ to get
\begin{equation}
\sum \exp\left( -(m +{k^2\over m}){\cal R}\right) 
= \int_0^{\infty}d\tau  K (x', x; \tau |\bar g) 
\exp\left( {-(m^2 + k^2)\tau}\right) 
\end{equation}
Differentiating this equation with respect to  $k^2$ gives:
\begin{equation}
\sum {\left({{\cal R}\over m} 
e^{-({k^2\over m}){\cal R}}\right) e^{-m{\cal R}}} 
= \int\limits^{\infty}_0 d \tau (\tau e^{-k^2 \tau}) 
e^{-m^2 \tau} K (x', x; \tau) 
\end{equation}
Fourier transforming on the variables $(k_1,k_2)$ with respect to two 
new variables $(l_1,l_2)$, we find
\begin{equation}
\sum  \int {d^2k \over \pi} {{\cal R}\over m}
e^{-m{\cal R}}\exp\left( {i {\bf k} . {\bf l} 
- {k^2\over m} {\cal R}}\right)    
= \int_0^{\infty} d\tau \left(\int {d^2k \over \pi} 
e^{i{\bf k}.{\bf l} - k^2 \tau} \right)\tau e^{-m^2 \tau} K 
\end{equation}
Or, equivalently,
\begin{equation}
\sum \exp (-m {\cal R} - {ml^2 \over 4{\cal R}}) 
= \int\limits^{\infty}_0 d\tau K 
\exp (-m^2 \tau - {l^2 \over 4\tau})  
\end{equation}
Defining $l^2=4L^2_p$, we get the final result
\begin{equation}
G_F(x,y|\bar g)
= \sum \exp -m ({\cal R} + {L^2 _p\over {\cal R}} ) 
= \int\limits^{\infty}_0 d \tau  K (x, x', \tau | \bar g)
\exp \left[ -(m^2\tau + {L^2 _p \over \tau})\right]\label{eqn:use}
\end{equation}
The above approach gives a surprisingly quick derivation of our 
result~(\ref{eqn:lastres}) provided we accept the definition of 
measure in (\ref{eqn:genform}) and set $L = 2L_p$. 
The analysis given above also generalizes to an arbitrary background 
with metric $\bar g_{ik}$.

Given the Kernel $K(x,y;\tau |\bar g)$ for a particle to propagate 
from $x$ to $y$ in proper time $\tau$ (in some background metric 
$\bar g_{ik}$), one would have originally evaluated the Feynman 
propagator by giving a weightage $\exp (-m^2\tau)$ and integrating 
over $\tau$. 
The effect of our modification is to change this weightage to $\exp(-m^2\tau-L^2_p/\tau)$. 
In deriving this result, we have not bothered to specify explicitly 
the measure in equation~(\ref{eqn:stpt}). 
To this extent, the derivation  is formal and not rigorous. 

\section{Conclusions}\label{sec:cnclsns}

One immediate consequence of this result is the interpretation in
terms of the `zero-point length' mentioned in the introduction.
We know that the Kernel $K(x,y;\tau|g)$ has an expansion of the form
\begin{equation}
K(x,y;\tau|\bar g) 
=\left( {1 \over 4 \pi \tau }\right)^{D/2} 
\exp\left(-{(x-y)^2 \over 4 \tau}\right)\left[1+.....\right]
\label{eqn:dew2}
\end{equation}
where the dots.....  represent metric-dependent corrections. 
Using equation~(\ref{eqn:dew2}) in equation~(\ref{eqn:use}) we can 
write our propagator as
\begin{equation}
G_F(x,y|\bar g)
=  \int\limits^{\infty}_0 d \tau e^{-m^2\tau} 
\left( {1 \over 4 \pi \tau }\right)^{D/2} 
\exp \left(-{(x-y)^2 +4 L^2_p\over 4 \tau}\right)[1 +....]
\end{equation}
Thus the net effect of our modification is to add a `zero-point length'
$4L^2_p$ to $(x-y)^2$ in the original propagator. 
In other words, {\it the modification of the path integral based on the 
principle of duality leads to results which are identical to adding a
`zero-point length' in the spacetime interval}. 

I wish to argue that the connection shown above is non-trivial; 
I know of no simple way of guessing this result. 
The standard Feynman propagator of quantum field theory can be obtained 
either through a lattice regularization of a path integral or from 
Schwinger's proper time representation. 
By adding a zero-point length in the Schwinger's representation we obtain 
a modified propagator. 
Alternatively, using the principle of duality, we could modify the 
expression for the path integral amplitude on the lattice and obtain---in 
the continuum limit---a modified propagator. 
Both these constructions are designed to  suppress energies larger than 
Planck energies. 
{\it However, there is absolutely no reason for these two expressions 
to be identical.} 
The fact that they are identical suggests that the principle of duality 
is connected in some deep manner with the spacetime intervals having a 
zero-point length.  
Alternatively, one may conjecture that any approach which introduces a 
minimum length scale in spacetime (like in string models) will lead to 
some kind of principle of duality. 
This conjecture seems to be true in conventional string theories though 
it must be noted that the term duality is used in somewhat different 
manner in string theories. 
(The concept  of duality in string theory is reviewed in several articles; 
see {\it e.g.}, refs.\cite{ashk94,giveon94,alvrz94,sch96,pol96,dne96} 
and the references cited therein. 
The closest to our approach seems to be the T-duality.)

The second obvious point, of course, is the improved ultraviolet behavior
in the theory. 
This has far reaching implications. 
For example, this ultraviolet finiteness  allows renormalization procedure
to be carried out without the need for regularization. 
Renormalized coupling constants now have no divergent pieces and will 
depend on the Planck length. 
In this sense, Planck length acts as a natural cut-off, as to be expected.

There are two other implications of this result which requires study. 
To begin with a Planck length cut-off is equivalent to changing the
density of states at high energies. 
The number of quantum states accessible to field theoretic system becomes
effectively finite. 
In the case of a black hole---for example---the number of microstates will 
be finite and will lead to a finite value for its entropy. 

The second issue is related to anomalies (like trace anomaly) in curved
spacetime. 
The conventional calculations does depend on the need to regularize the
expressions in one way or the other~\cite{parker79}. 
With ultraviolet finiteness it is not clear whether the anomalies will 
survive or not. 
(Some preliminary calculations suggest that the anomalies will vanish.) 
These and related issues are under investigation.

\appendix
\section{Evaluation of the sum}\label{app:sumeval}

We need to evaluate the sum
\begin{equation}
S(a, b) 
\equiv \sum\limits^{\infty}_{n=1} e^{-a^2n - {b^2 \over n}} 
= \sum\limits^{\infty}_{n=0}e^{-a^2n - {b^2 \over n}} 
\qquad (b\not= 0) 
\end{equation}
To do this, we introduce two real variables $(x, y)$ and 
write $b^2 \equiv (x^2 + y^2 )/4$.  
Then we have the identity:
\begin{equation}
\exp(-{b^2 \over n})
= {\sqrt n \over \sqrt \pi} 
\int_{-\infty}^{\infty}dk_x 
e^{-n k^2_x + i k_x x}. {\sqrt n \over \sqrt \pi}
\int_{-\infty}^{\infty}dk_y e^{-n k^2_y + i k_y y}
=\int {d^2 {\bf k} \over \pi} ne^{- nk^2 + i {\bf k} . {\bf x} }
\end{equation}
So the sum we need is,
\begin{equation}
 S (a, {\bf x} ) 
= \int {d^2 {\bf k} \over \pi} e^{i{\bf k} . {\bf x}} 
\sum\limits^{\infty}_{n=1} ne^{-n (a^2 + {\bf k}^2 )}; 
\qquad|{\bf x}|= 2b
\end{equation}
with ${\bf x} = (x, y)$ being two-dimensional vector. 
Now
\begin{equation}
\sum\limits^{\infty}_{n=0}n e^{- \mu n} 
= - {\partial \over \partial \mu} 
\left( {1 \over 1 - e^{-\mu} }\right) = {e^{-\mu} \over (1 -e^{-\mu})^2 }
\end{equation}
giving 
\begin{eqnarray}
S(a, {\bf x})
&=&\int{d^2 {\bf k} \over \pi} e^{i{\bf k} . {\bf x}} 
{e^{-(a^2 + k^2)} \over  (1 -e^{-(a^2 + k^2)})^2 } 
;\quad   
|{\bf x} | = 2b \nonumber\\
&= &\int_{0}^{2 \pi} {d \theta \over \pi} 
\int_{0}^{\infty} kdk e^{2ikb\cos \theta} 
{e^{-(a^2 + k^2)} \over  [1 -e^{-(a^2 + k^2)}]^2} 
\end{eqnarray}
To do the $\theta$-integral, we need the result:
\begin{equation}
I = \int_{0}^{2 \pi} d \theta e^{i \mu \cos \theta} = 2 \pi J_0 (\mu ) 
\end{equation}

Using this we get
\begin{equation}
S(a, b)
=2 \int_0^{\infty} k dk J_0 (2kb) 
{e^{-(a^2 + k^2)} \over  [1 -e^{- (a^2 + k^2 )]^2}}
=\int_0^{\infty}{q dq \over 2b^2} 
{J_0(q) e^{-(a^2 + {q^2 \over 4b^2})}
\over[1 -e^{-(a^2+ {q^2 \over 4b^2})}]^2}
\end{equation}
This is the result quoted in the text.

\end{document}